\documentclass[letterpaper]{article} \usepackage{aaai2026}
\usepackage{times}  \usepackage{helvet}  \usepackage{courier}  \usepackage[hyphens]{url}  \usepackage{graphicx}    \usepackage{natbib}  \usepackage{caption}     \usepackage{algorithm}
\usepackage{algorithmic}
\usepackage{amsmath}
\usepackage{adjustbox}
\usepackage{xcolor}
\usepackage{soul}
\usepackage{booktabs}
\usepackage{pifont}
\usepackage{newfloat}
\usepackage{listings}
\usepackage{xspace}
\usepackage{fontawesome5}
\usepackage{tikz}
\usetikzlibrary{arrows.meta,positioning,calc,shapes.geometric,shapes.misc}

\DeclareCaptionStyle{ruled}{labelfont=normalfont,labelsep=colon,strut=off} 
\floatstyle{ruled}
\newfloat{listing}{tb}{lst}{}
\floatname{listing}{Listing}

\usepackage[inline]{enumitem}

\title{The Big Ban Theory:\\A Pre- and Post-Intervention Dataset of Online Content Moderation Actions\thanks{\textcolor{red}{Article published in \textit{ICWSM'26 – 20th AAAI Conference on Web and Social Media}. DOI: https://doi.org/10.1609/icwsm.v20i1.42780. Please, cite the published version.}}}

\author {
    Aldo Cerulli\textsuperscript{\rm 1,2}\thanks{Equal contributions.},
    Lorenzo Cima\textsuperscript{\rm 1,3}\footnotemark[1],
    Benedetta Tessa\textsuperscript{\rm 1,2}\thanks{Corresponding author.},
    Serena Tardelli\textsuperscript{\rm 1},
    Stefano Cresci\textsuperscript{\rm 1}
}
\affiliations {
    \textsuperscript{\rm 1}IIT-CNR, Italy\\
    \textsuperscript{\rm 2}Dept. of Computer Science, University of Pisa, Italy\\
    \textsuperscript{\rm 3}Dept. of Information Engineering, University of Pisa, Italy\\
   \{name.surname\}@iit.cnr.it
}

\newcommand{\data}{\texttt{TBBT}\xspace}
\newcommand{\subr}[1]{{\small\texttt{r/#1}}}
\newcommand{\cmark}{\ding{51}} \newcommand{\xmark}{\textcolor{gray!80}{\ding{55}}} 
\begin{document}

\maketitle

\begin{abstract}
Online platforms rely on moderation interventions to curb harmful behavior such as hate speech, toxicity, and the spread of mis- and disinformation. Yet research on the effects and possible biases of such interventions faces multiple limitations. For example, existing works frequently focus on single or a few interventions, due to the absence of comprehensive datasets. As a result, researchers must typically collect the necessary data for each new study, which limits opportunities for systematic comparisons. To overcome these challenges, we introduce \textit{The Big Ban Theory} (\data)---a large dataset of moderation interventions. \data covers 25 interventions of varying type, severity, and scope, comprising in total over 339K users and nearly 39M posted messages on Reddit and Voat. For each intervention, we provide standardized metadata and pseudonymized user activity collected three months before and after its enforcement, enabling consistent and comparable analyses of intervention effects. In addition, we provide a descriptive exploratory analysis of the dataset, along with several use cases of how it can support research on content moderation. With this dataset, we aim to support researchers studying the effects of moderation interventions and to promote more systematic, reproducible, and comparable research. \end{abstract}
\begin{links}
\link{Datasets}{https://doi.org/10.5281/zenodo.18245669}
\end{links}

\section{Introduction}
As online platforms grow in scale and societal relevance, they increasingly rely on a wide range of moderation interventions to curb illegal, harmful, or otherwise problematic behavior~\cite{grimmelmann2015virtues,gillespie2018custodians}. The volume, diversity, and frequency of such interventions have expanded substantially~\cite{shahi2025year}, reflecting both the growth of online user activity and the complexity of governing large-scale socio-technical systems. As a result, the current landscape of content moderation is highly multifaceted, encompassing multiple platforms, heterogeneous intervention types, and a broad spectrum of potential outcomes.

Frequently applied interventions include \textit{hard} actions, such as user or community bans~\cite{jhaver2021evaluating,cima2024great} and content removals~\cite{jhaver2024bystanders}, as well as an expanding set of \textit{soft} interventions encompassing warning messages~\cite{papakyriakopoulos2022impact,ling2023learn}, demonetization~\cite{caplan2020tiered}, counterspeech~\cite{garland2022impact,cima2025contextualized}, and other forms of friction or visibility reduction~\cite{chandrasekharan2022quarantined,jahn2025perspective}. These interventions are typically applied as discrete, time-stamped actions that modify ongoing patterns of user and community behavior, thereby creating natural---though methodologically challenging---opportunities to measure their effects by comparing behavior before and after the moderation takes place.

Within this complex and rapidly evolving context, there is a strong need for empirical studies that assess the strategies, consequences, and broader implications of content moderation. Relevant tasks include evaluating whether moderation practices are fair, transparent, and effective. Concerning effectiveness, prior research has reported mixed findings~\cite{cima2025investigating}. For example, some interventions appear to achieve their intended goals~\cite{costello2024durably,horta2025deplatforming}, others show limited or no measurable impact~\cite{aslett2022news}, and some are associated with unintended or counterproductive effects~\cite{bail2018exposure,pennycook2020implied,trujillo2022make}. Adverse effects may also include behavioral adaptation, evasion, or migration, whereby affected users relocate their activity to other spaces---either on the same platform or elsewhere entirely---leading to outcomes that manifest outside the directly moderated space~\cite{horta2021platform}. Amid this scenario, resources that enable systematic, comparative, and reusable research on the application and effects of moderation interventions are increasingly needed to support understanding of how moderation operates and with what consequences.

Nonetheless, existing resources are overall limited and fragmented. The ``post-API'' era has been characterized by a significant reduction in access to platform data, and subsequent efforts to mitigate these constraints have yet to restore broad and reliable access~\cite{jaursch2024dsa,mimizuka2025post}. This creates clear obstacles for research on online spaces.As a result, scholars often struggle to collect and curate adequate datasets. Moreover, existing datasets of moderation interventions are typically narrow in scope, covering one or a few interventions, usually on a single platform~\cite{trujillo2022make,shen2022tale,russo2023spillover}. The remaining datasets are mostly general-purpose and were not designed to capture moderation interventions as analyzable events~\cite{baumgartner2020pushshift,seckin2025labeled}, making them difficult to use directly and systematically for studying the effects of moderation actions. Finally, existing datasets have been collected using heterogeneous definitions, methodologies, and parameters, which limits their interoperability and prevents comprehensive, systematic analyses. On the flip side, some recent initiatives have partially improved data availability. X has made publicly available comprehensive datasets related to its \textit{Community Notes} program,\footnote{\url{https://communitynotes.x.com/guide/en/under-the-hood/download-data}} including note-level information and associated metadata~\cite{chuai2024did,slaughter2025community}. Furthermore, the Digital Services Act has established the Transparency Database,\footnote{\url{https://transparency.dsa.ec.europa.eu/}} a centralized open repository where large platforms must report moderation actions taken in the EU. However, while valuable, this resource is limited to \textit{metadata} about moderation decisions---such as the type of violation or intervention---and does not include the moderated content, conversational context, or pre- and post-intervention data~\cite{kaushal2024automated,trujillo2025dsa}. Consequently, it cannot support analyses of moderation effects, biases, or other downstream tasks. In the absence of standardized, intervention-centered datasets, researchers must therefore reconstruct moderation events and assemble relevant data manually from disparate sources. 

\textbf{Contribution.} We contribute to addressing these limitations by introducing \textit{The Big Ban Theory} (\data) dataset---a large-scale, intervention-centered collection of moderation data. The dataset comprises 25 moderation interventions of multiple types---such as bans, quarantines, and content removals---enforced between 2015 and 2023. All interventions occurred on Reddit, but some caused user migrations to Voat, for which we also provide data. Across all interventions, \data contains pseudonymized data about 38,700,732 messages published by 339,125 distinct users. 
The dataset includes systematically aligned observations from fixed windows before and after each intervention, allowing to examine both the context preceding moderation---such as behavioral or content-level signals that may trigger enforcement---and its subsequent effects, including changes in user behavior, participation patterns, and community dynamics. \data is designed to support a broad range of analyses that are time-demanding and challenging with existing resources. These include assessing the fairness, consistency, and heterogeneity of moderation practices; studying short- and medium-term responses to different intervention types; and examining potential spillover effects. 
While not exhaustive, \data provides a structured foundation for comparative and cumulative research on moderation interventions, supporting further work and the development of shared, reusable resources. \section{Related Work}

\subsection{Content Moderation Datasets}
Current works on online content moderation typically rely on datasets comprising users' posts and comments, together with basic metadata such as text, creation and deletion timestamps, user identifiers, and community membership. Thanks to this information, they make it possible to understand how users react to content moderation interventions by comparing data collected before and after the intervention.
For example, studies on community interventions such as bans or quarantines on Reddit analyzed longitudinal datasets of user comments to track changes in a variety of indicators, including activity, toxicity, in-group vocabulary, factual reporting, political bias, and community participation \cite{cima2024great,cima2025investigating,trujillo2021echo,trujillo2022make,shen2022tale,trujillo2023one}. By assessing the outcome of the intervention at both individual and community levels, these studies were able to measure how these indicators evolved over time.
Beyond Reddit, research on other platforms---such as X and Wikipedia---similarly relies on longitudinal datasets built from posts made by users to study the effects of deplatforming, quarantine and other types of interventions \cite{horta2025deplatforming,zannettou2021won,jhaver2021evaluating}.

Taken together, this body of work highlights the central role of datasets that combine basic account- and activity-level information with temporal alignment around a moderation event. While such datasets have enabled valuable case-specific analyses of moderation outcomes, they are typically constructed for a single intervention or a small set of related interventions on a single platform. As a result, findings are often difficult to compare across studies, and it remains challenging to generalize conclusions across different types of moderation actions, platforms, or contexts. This limitation underscores the need for more systematic, reusable resources that support comparative analyses of moderation interventions across settings.

\subsection{Effects of Moderation Interventions}
A large body of prior work relies on post-hoc analyses to assess how users, content, and communities respond to moderation interventions. These studies typically adopt descriptive or quasi-causal designs that compare indicators extracted from posts or comments---such as activity, toxicity, linguistic features, or participation patterns---before and after an intervention takes place \cite{trujillo2021echo,cima2024great,shen2022tale,copland2020reddit}. In their simplest form, such analyses measure differences across pre- and post-intervention windows and assess their statistical significance using techniques such as ANOVA or related tests~\cite{trujillo2023one}. More rigorous approaches employ causal inference methods, including Difference-in-Differences, to account for temporal trends and unobserved confounders~\cite{cima2025investigating}. When more granular temporal data are available, researchers often aggregate outcomes into daily or weekly time series and apply methods such as interrupted time series analysis or Bayesian structural time series models to estimate both immediate and longer-term intervention effects \cite{trujillo2022make,horta2023automated,kumarswamy2023impact}. While much of this work focuses on user- or community-level interventions, similar post-hoc designs have also been applied to content-level moderation, such as warning labels on disputed posts, to study their effects on perceived credibility and user engagement, sometimes revealing unintended or counterintuitive outcomes \cite{zannettou2021won}.

Beyond descriptive assessment, the outcomes of such post-hoc analyses are often used to derive ground-truth labels that enable predictive approaches. In this setting, intervention effects estimated from pre- and post-intervention data serve as targets for models trained solely on pre-intervention signals, with the goal of anticipating how users, content, or communities will respond to future moderation actions. Prior work has used this approach to predict user-level responses, including abandonment~\cite{tessa2025beyond} or evasion following bans~\cite{niverthi2022characterizing}, as well as changes in activity, toxicity, and participation diversity~\cite{tessa2025quantifying}. Other studies operate at the community level, aiming to identify communities likely to face moderation actions based on structural or behavioral indicators \cite{habib2022proactive}. Complementary work focuses on content-level prediction, leveraging metadata such as posting context, author history, and temporal features to forecast whether posts will later be moderated \cite{kurdi2020video,chandrasekharan2019crossmod,paudel2023lambretta}.

Overall, these works illustrate how intervention-centered datasets not only support retrospective evaluation, but also underpin predictive models that seek to anticipate moderation outcomes and inform proactive strategies.
 \begin{figure}[t]
    \centering
    \resizebox{\columnwidth}{!}{\begin{tikzpicture}[
  font=\large,
  >=Latex,
  TimeAxisWidth/.store in=\TimeAxisWidth, TimeAxisWidth=11, YCenterInBand/.store in=\YCenterInBand, YCenterInBand=4, YCenterOutBand/.store in=\YCenterOutBand, YCenterOutBand=1.5, BandHalfHeight/.store in=\BandHalfHeight, BandHalfHeight=0.45, XTimeStart/.store in=\XTimeStart, XTimeStart=0.0, XInterventionTime/.store in=\XInterventionTime, XInterventionTime=5.0, XTimeEnd/.store in=\XTimeEnd, XTimeEnd=\TimeAxisWidth, XDataEnd/.store in=\XDataEnd, XDataEnd=\TimeAxisWidth - 1, slice/.style={draw, thick, rounded corners=5pt, minimum height=7mm, align=center},
  inslice/.style={slice, fill=black!3}, outslice/.style={slice, fill=black!3}, area/.style={draw, gray, fill=black!3, rounded corners=0pt, inner sep=2mm},
  event/.style={draw, fill=white, align=center, inner sep=1.5mm},
  lab/.style={align=left}
]
\definecolor{inbef}{HTML}{fdbf6f}
\definecolor{inaft}{HTML}{cab2d6}
\definecolor{outbef}{HTML}{b2df8a}
\definecolor{outaft}{HTML}{a6cee3}
\definecolor{moderation}{HTML}{e31a1c}

\draw[thick,-{Latex}] (\XTimeStart,0) -- (\XTimeEnd,0) node[below=1.2mm] {time};
\draw[thick, color=moderation] (\XInterventionTime,-0.12) -- (\XInterventionTime,0.12);
\node[below=1.2mm, color=moderation] at (\XInterventionTime,0) {$t_0$};
\draw[thick] (\XTimeStart,-0.12) -- (\XTimeStart,0.12);
\node[below=1.2mm] at (\XTimeStart,0) {$t_{\text{before}}$};
\draw[thick] (\XDataEnd,-0.12) -- (\XDataEnd,0.12);
\node[below=1.2mm] at (\XDataEnd,0) {$t_{\text{after}}$};

\draw[area] (\XTimeStart, \YCenterInBand-\BandHalfHeight-0.75) rectangle (\XDataEnd, \YCenterInBand+\BandHalfHeight+0.75);
\node[anchor=west] at (\XTimeStart+0.15, \YCenterInBand+\BandHalfHeight+0.35) {\texttt{IN}: moderated space};
\draw[area] (\XTimeStart, \YCenterOutBand-\BandHalfHeight-0.75) rectangle (\XDataEnd, \YCenterOutBand+\BandHalfHeight+0.15);

\draw[inslice, fill=inbef] (\XTimeStart, \YCenterInBand-\BandHalfHeight) rectangle (\XInterventionTime, \YCenterInBand+\BandHalfHeight);
\node at ({(\XTimeStart+\XInterventionTime)/2}, \YCenterInBand) {\texttt{IN-BEFORE}};
\draw[inslice, fill=inaft, dashed] (\XInterventionTime, \YCenterInBand-\BandHalfHeight) rectangle (\XDataEnd, \YCenterInBand+\BandHalfHeight);
\node[color=black!3] at ({(\XInterventionTime+\XDataEnd)/2}, \YCenterInBand) {\texttt{IN-AFTER}};

\draw[outslice, fill=outbef] (\XTimeStart, \YCenterOutBand-\BandHalfHeight) rectangle (\XInterventionTime, \YCenterOutBand+\BandHalfHeight);
\node at ({(\XTimeStart+\XInterventionTime)/2}, \YCenterOutBand) {\texttt{OUT-BEFORE}};
\draw[outslice, fill=outaft] (\XInterventionTime, \YCenterOutBand-\BandHalfHeight) rectangle (\XDataEnd, \YCenterOutBand+\BandHalfHeight);
\node at ({(\XInterventionTime+\XDataEnd)/2}, \YCenterOutBand) {\texttt{OUT-AFTER}};

\node[anchor=center] (users) at (\XTimeStart+2.5,\YCenterInBand-\BandHalfHeight-0.35) {{\Large\faUsers} \textbf{affected users}};
\draw[->, thick] (users.south) -- (2.5, \YCenterOutBand+\BandHalfHeight+0.25) node[midway, fill=white, inner sep=1.2pt, outer sep=0pt] {\small\faComments};

\draw[thick, dashed, color=moderation] (\XInterventionTime, 0) -- (\XInterventionTime, 5.5);
\fill[moderation] (\XInterventionTime, 5.5) circle (3.5pt);
\node[anchor=south, inner sep=0pt] at (\XInterventionTime, 5.75) {\textbf{moderation intervention}};

\node[thick, draw=white, fill=moderation, regular polygon, regular polygon sides=3, minimum size=8mm, rounded corners=2pt] at (\XInterventionTime, \YCenterInBand-0.07) {};

\node at (\XInterventionTime, \YCenterInBand) {\textcolor{white}{\small\faExclamation}};

\node[anchor=west, fill=black!3, inner sep=1.5pt, outer sep=0pt] at (\XTimeStart+0.15, \YCenterOutBand+\BandHalfHeight-1.35) {\texttt{OUT}: other spaces in which \textbf{affected users} participate};

\end{tikzpicture}
 }
    \caption{For an intervention occurring at time $t_0$, we identify four data slices: \texttt{IN-BEFORE} and \texttt{IN-AFTER} (if available) capture activity in the moderated space before and after the intervention, while \texttt{OUT-BEFORE} and \texttt{OUT-AFTER} capture activity by the same users outside that space.} 
    \label{fig:data-model}
\end{figure}
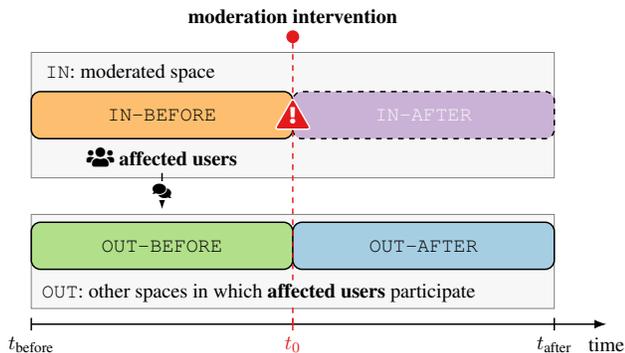

\section{Methodology}
\subsection{Data Model}
We model content moderation as a discrete, time-stamped intervention that affects a set of users in a moderated space (e.g., a community or platform). Figure~\ref{fig:data-model} provides an overview of the data model. Accordingly, for a given intervention, we identify a cohort of affected users and partition their activity along two orthogonal dimensions: \textit{(i)} \textbf{time} (before \textit{vs.} after the intervention) and \textit{(ii)} \textbf{space} (inside \textit{vs.} outside the moderated space). This results in four possible data slices: \texttt{IN-BEFORE}, \texttt{IN-AFTER}, \texttt{OUT-BEFORE}, and \texttt{OUT-AFTER}. The two \texttt{IN} slices capture activity within the space where moderation occurred, while the two \texttt{OUT} slices capture activity by the same users elsewhere (e.g., another part of the same platform, or a different platform altogether). A key motivation for this design is that, for some interventions (e.g., community bans), the \texttt{IN-AFTER} slice is structurally unavailable, as the moderated space no longer exists after the intervention. In such cases, incorporating \texttt{OUT} activity before and after the intervention still enables the analysis of post-intervention and downstream effects.

\subsection{Selection of Moderation Interventions}
We conducted a targeted review of prior work on content moderation, focusing on interventions that have been repeatedly analyzed or highlighted as particularly relevant in the existing literature. For each study, we recorded the type of analyzed intervention and evaluated the feasibility of collecting the corresponding data under our data model. A limited number of more recent cases were additionally identified through publicly available resources, such as the subreddit \texttt{r/reclassified}\footnote{\url{https://www.reddit.com/r/reclassified/}}
, which tracks newly imposed moderation actions on Reddit.
Based on this process, we selected a longitudinal set of 25 moderation interventions carried out between 2015 and 2023, spanning two platforms---Reddit and Voat---and covering multiple intervention types, including community bans, quarantines, and post removals. The set of considered interventions is reported in the leftmost half of Table~\ref{tab:dataset}. This selection is constrained by the existing literature on moderation interventions and by data availability, and we acknowledge that it is neither exhaustive nor uniformly distributed across time, platforms, or intervention types. Nonetheless, it provides a grounded and extensive resource for downstream analyses, while leaving room for expansion to additional interventions and platforms as future work.

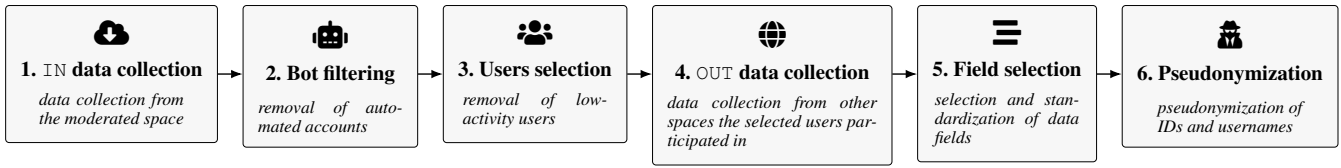
\begin{figure*}[t]
    \centering
    \resizebox{\textwidth}{!}{\begin{tikzpicture}[
  font=\small,
  >=Latex,
  stepbox/.style={draw, rounded corners=2pt, fill=black!3, align=center, inner sep=3mm, minimum height=18mm, anchor=north west},
  steptitle/.style={font=\bfseries},
  stepdesc/.style={font=\scriptsize, text=black!70},
  arrow/.style={->, thick},
  boxw/.store in=\boxw, boxw=2.95cm,
  gap/.store in=\gap, gap=5mm,
  descw/.store in=\descw, descw=2.90cm,
  descww/.store in=\descww, descww=4.20cm
]

\node[stepbox, minimum width=\boxw] (s1) at (0,0) {{\fontsize{16}{20}\selectfont\faCloudDownload*}\\[3mm] {\Large\textbf{1. }\texttt{IN} \textbf{data collection}}\\[3mm] \parbox{\descw}{\normalsize\textit{data collection from the moderated space}}};

\node[stepbox, minimum width=\boxw] (s2) at ($(s1.north east)+(\gap,0)$) {{\fontsize{16}{20}\selectfont\faRobot}\\[3mm] {\Large\textbf{2. Bot filtering}}\\[3mm] \parbox{\descw}{\normalsize\textit{removal of automated accounts}}};

\node[stepbox, minimum width=\boxw] (s3) at ($(s2.north east)+(\gap,0)$) {{\fontsize{16}{20}\selectfont\faUsers}\\[3mm] {\Large\textbf{3. Users selection}}\\[3mm] \parbox{\descw}{\normalsize\textit{removal of low-activity users}}};

\node[stepbox, minimum width=\boxw] (s4) at ($(s3.north east)+(\gap,0)$) {{\fontsize{16}{20}\selectfont\faGlobe}\\[3mm] {\Large\textbf{4. }\texttt{OUT} \textbf{data collection}}\\[3mm] \parbox{\descww}{\normalsize\textit{data collection from other spaces the selected users participated in}}};

\node[stepbox, minimum width=\boxw] (s5) at ($(s4.north east)+(\gap,0)$) {{\fontsize{16}{20}\selectfont\faStream}\\[3mm] {\Large\textbf{5. Field selection}}\\[3mm] \parbox{\descw}{\normalsize\textit{selection and standardization of data fields}}};

\node[stepbox, minimum width=\boxw] (s6) at ($(s5.north east)+(\gap,0)$) {{\fontsize{16}{20}\selectfont\faUserSecret}\\[3mm] {\Large\textbf{6. Pseudonymization}}\\[3mm] \parbox{\descw}{\normalsize\textit{pseudonymization of IDs and usernames}}};

\path (s1.north west) -- (s1.south west) coordinate[pos=0.5] (yMid);
\draw[arrow] ($(s1.east |- yMid)$) -- ($(s2.west |- yMid)$);
\draw[arrow] ($(s2.east |- yMid)$) -- ($(s3.west |- yMid)$);
\draw[arrow] ($(s3.east |- yMid)$) -- ($(s4.west |- yMid)$);
\draw[arrow] ($(s4.east |- yMid)$) -- ($(s5.west |- yMid)$);
\draw[arrow] ($(s5.east |- yMid)$) -- ($(s6.west |- yMid)$);

\end{tikzpicture}
 }
    \caption{Overview of the six steps of the data collection and processing pipeline used to construct the dataset.}
    \label{fig:data-preparation}
\end{figure*}

\subsection{Data Collection and Preparation}
For each selected intervention, we applied the following standardized pipeline to collect and prepare the data, also outlined in Figure~\ref{fig:data-preparation}:
\begin{enumerate}
    \item \texttt{IN} \textbf{data collection.} Based on the intervention timestamp $t_0$, we retrieved raw data from the moderated space covering a fixed window of three months before and after the intervention, when available. With reference to Figure~\ref{fig:data-model}, this means that $t_{\text{after}},t_{\text{before}}=t_0\pm3\text{ months}$. The data was collected from publicly available resources. Specifically, Reddit data was collected from a torrent file,\footnote{\url{https://academictorrents.com/details/9c263fc85366c1ef8f5bb9da0203f4c8c8db75f4}} while Voat data from a public Zenodo repository by~\citet{mekacher2022can}, with CC BY 4.0 (redistributable) license.\footnote{\url{https://zenodo.org/records/5841668}}
    \item \textbf{Bot filtering.} Following prior work, we removed bot accounts by excluding users who posted two or more comments with identical timestamps~\cite{hurtado2019bot,cima2025investigating}. We further filtered out self-declared bots and conducted a manual validation of the bot filtering process.
    \item \textbf{Selection of affected users.} We retained only users who posted at least ten messages in the moderated space in the three months preceding the intervention. This step filters out low-activity accounts and focuses the dataset on users meaningfully affected by the moderation action.
    \item \texttt{OUT} \textbf{data collection.} In cases when the \texttt{IN-AFTER} data slice is unavailable, we collected activity by the affected users outside the moderated space, such as in other communities on the same platform or on alternative platforms to which those users migrated.
    \item \textbf{Field selection and standardization.} From the available raw data, we selected a subset of fields based on relevance and consistency. Because \data spans multiple years and platforms with evolving data schemas, this step harmonizes field names and formats, providing longitudinal consistency and alignment across Reddit and Voat, to the extent possible. \data field names follow Reddit's naming conventions, since it accounts for the majority of the dataset. Appendix Table~\ref{tab:attributes-description} reports the full list and description of provided fields.
    \item \textbf{Pseudonymization.} To protect user privacy, we de-identified \data by pseudonymizing all identifiers via a deterministic cryptographic hash applied to all IDs and usernames, both in the metadata and message text. This step preserves internal consistency while mitigating re-identification risks.
\end{enumerate}

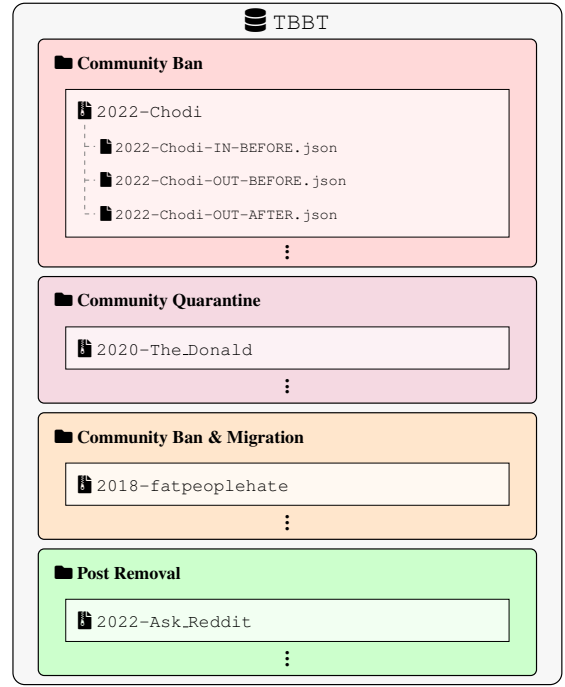
\begin{figure}[t]
    \centering
    \resizebox{0.875\columnwidth}{!}{\begin{tikzpicture}[every node/.style={font=\sffamily}, node distance=6mm]
\tikzset{
  tbbt/.style={draw, rounded corners=8pt, thick, fill=black!3, minimum width=13cm, minimum height=16.15cm},
  section/.style={draw, rounded corners=4pt, thick, inner sep=10pt, minimum width=11.8cm, minimum height=5.4cm},
  section-small/.style={draw, rounded corners=4pt, thick, inner sep=10pt, minimum width=11.8cm, minimum height=3cm},
  dataset/.style={draw, rounded corners=0pt, inner sep=10pt, minimum width=10.5cm, minimum height=3.5cm},
  dataset-small/.style={draw, rounded corners=0pt, inner sep=10pt, minimum width=10.5cm, minimum height=1cm},
  title/.style={align=left, anchor=north west, yshift=-8pt, xshift=8pt},
  title-intervention/.style={align=left, anchor=north west, yshift=-5pt, xshift=5pt}
}
\node[tbbt] (tbbt) {};
\node[yshift=-12pt] at (tbbt.north) {\fontsize{16}{20}\selectfont\faDatabase~\texttt{TBBT}};

\node[section, fill=red!15, anchor=north, below=0.85cm of tbbt.north] (ban) {};
\node[title, font=\bfseries] at (ban.north west) {\Large\faFolder~community-ban};
\node[dataset, fill=red!5, below=12mm of ban.north] (chodi) {};
\node[title-intervention] (chodi-title) at (chodi.north west) {\Large\faFileArchive~\texttt{2022-chodi}};
\coordinate (leftedge) at ([xshift=20pt]chodi.west);
\node[font=\normalsize, anchor=west] (chodi-in-bef) at ([yshift=4mm]leftedge) {\faFile~\texttt{2022-chodi-IN-BEFORE.json}};
\node[font=\normalsize, anchor=west] (chodi-out-bef) at ([yshift=-4mm]leftedge) {\faFile~\texttt{2022-chodi-OUT-BEFORE.json}};
\node[font=\normalsize, anchor=west] (chodi-out-aft) at ([yshift=-12mm]leftedge) {\faFile~\texttt{2022-chodi-OUT-AFTER.json}};
\coordinate (spine-start) at ([xshift=13pt]chodi.west);
\coordinate (spine-end) at ([xshift=13pt]chodi-out-aft);
\draw[dashed, gray] (spine-start |- chodi-title.south) -- (spine-start |- spine-end);
\draw[dashed, gray] (spine-start |- chodi-in-bef) -- (chodi-in-bef.west);
\draw[dashed, gray] (spine-start |- chodi-out-bef) -- (chodi-out-bef.west);
\draw[dashed, gray] (spine-start |- chodi-out-aft) -- (chodi-out-aft.west);
\node[font=\Huge] at ([yshift=-20mm] chodi.center) {\vdots};

\node[section-small, fill=purple!15, below=0.2cm of ban] (qua) {};
\node[title, font=\bfseries] at (qua.north west) {\Large\faFolder~community-quarantine};
\node[dataset-small, fill=purple!5, below=12mm of qua.north] (donald) {};
\node[title-intervention] at (donald.north west) {\Large\faFileArchive~\texttt{2020-the\_donald}};
\node[font=\Huge] at ([yshift=-8mm] donald.center) {\vdots};

\node[section-small, fill=orange!20, below=0.2cm of qua] (mig) {};
\node[title, font=\bfseries] at (mig.north west) {\Large\faFolder~community-ban-migration};
\node[dataset-small, fill=orange!5, below=12mm of mig.north] (gw) {};
\node[title-intervention] at (gw.north west) {\Large\faFileArchive~\texttt{2018-fatpeoplehate}};
\node[font=\Huge] at ([yshift=-8mm] gw.center) {\vdots};

\node[section-small, fill=green!20, below=0.2cm of mig] (rem) {};
\node[title, font=\bfseries] at (rem.north west) {\Large\faFolder~post-removal};
\node[dataset-small, fill=green!5, below=12mm of rem.north] (ask) {};
\node[title-intervention] at (ask.north west) {\Large\faFileArchive~\texttt{2022-ask\_reddit}};
\node[font=\Huge] at ([yshift=-8mm] ask.center) {\vdots};

\end{tikzpicture}
 }
    \caption{Dataset structure, showing the top-level folders by intervention type, the intervention-level compressed archives, and the standardized JSON files corresponding to the available data slices for each intervention.}
    \label{fig:structure}
\end{figure}

\subsection{Data Structure, Access, and FAIR Principles}
\data is structured to mirror the intervention-centered data model described earlier. The overall dataset structure is outlined in Figure~\ref{fig:structure}. At the top level, it comprises four folders, one per intervention type. Within each folder, data are further organized into intervention-level subfolders, compressed for efficient storage and access. Each of these compressed subfolders corresponds to a single moderation intervention (i.e., one row of Table~\ref{tab:dataset}). 
Each intervention-level subfolder contains one standardized, line-delimited JSON file for each slice of data available for that intervention, as per Table~\ref{tab:dataset}. In particular, the \texttt{IN-BEFORE} file is always available, while the \texttt{IN-AFTER}, \texttt{OUT-BEFORE}, and \texttt{OUT-AFTER} are provided optionally depending on the type of intervention and data availability. Each line in these files represents a single comment encoded as a JSON object with its associated attributes.
The attributes included in \data are a subset of those provided by the Python Reddit API Wrapper (PRAW).\footnote{\url{https://praw.readthedocs.io/en/stable/}} \data fully adheres to the FAIR\footnote{\url{https://force11.org/info/the-fair-data-principles/}} principles:
\begin{itemize}
    \item \textbf{Findable}: The dataset is publicly available on Zenodo with a persistent DOI:\\ \url{https://doi.org/10.5281/zenodo.18245669}.
    \item \textbf{Accessible}: The dataset is freely accessible via a recognized repository with clear licensing agreement that ensures long-term availability and responsible use.
    \item \textbf{Interoperable}: All data are provided in a standardized JSON format, ensuring compatibility and integration with other tools and datasets.
    \item \textbf{Re-usable}: The dataset comes with comprehensive documentation that supports future research.
\end{itemize}
 \begin{table*}[t!]
\centering
\setlength{\tabcolsep}{4pt}
\adjustbox{max width=\textwidth}{
\begin{tabular}{rrlll  rr rr rr rr}
\toprule
&&&&& \multicolumn{8}{c}{\textit{provided data}} \\
\cmidrule(lr){6-13}
\multicolumn{5}{c}{\textit{moderation intervention}} &
\multicolumn{2}{c}{\texttt{IN-BEFORE}} &
\multicolumn{2}{c}{\texttt{IN-AFTER}} &
\multicolumn{2}{c}{\texttt{OUT-BEFORE}} &
\multicolumn{2}{c}{\texttt{OUT-AFTER}} \\
\cmidrule(lr){1-5} \cmidrule(lr){6-7} \cmidrule(lr){8-9} \cmidrule(lr){10-11} \cmidrule(lr){12-13}
\textbf{date} $t_0$ & \textbf{target} & \textbf{type} & \textbf{moderated space} & \textbf{platform(s)}  & 
\multicolumn{1}{c}{\faUserFriends~$\downarrow$} & \multicolumn{1}{c}{\faComments} &
\multicolumn{1}{c}{\faUserFriends} & \multicolumn{1}{c}{\faComments} &
\multicolumn{1}{c}{\faUserFriends} & \multicolumn{1}{c}{\faComments} &
\multicolumn{1}{c}{\faUserFriends} & \multicolumn{1}{c}{\faComments} \\
\midrule
1-30 Jun 22 & post & removal& \subr{Ask\_Reddit} & Reddit 
& 273,792 & 11,661,920 & 217,978  & 7,343,707 & --  & --  & --  & --  \\
26 Jun 19 & community & quarantine & \subr{The\_Donald}  & Reddit
& 22,410 & 1,939,536 & 17,134 & 1,526,916 & --  & --  & --  & --  \\
6 Aug 19 & community & quarantine  & \subr{ChapoTrapHouse} & Reddit
& 11,922 & 866,206 & 8,855 & 601,161 & --  & --  & --  & --  \\
28 Sep 18 & community & quarantine  & \subr{Braincels} & Reddit
& 7,414 & 1,116,210  & 2,397& 385,258 & --  & --  & --  & --    \\
29 Jun 20 & community & ban  & \subr{ChapoTrapHouse} & Reddit
& 7,153 & 865,937 & \xmark\  & \xmark\  & 6,743 & 1,272,364 & 6,324 & 1,279,975 \\
1-30 Jun 22 & post & removal& \subr{science}  & Reddit 
& 5,028 & 100,346 & 3,749 & 48,236 & --  & --  & --  & --  \\
12 Sep 18 & community & ban  & \subr{greatawakening} & Reddit
& 4,550 & 329,183 & \xmark\  & \xmark\  & 3,570 & 690,552 & 2,956 & 620,307 \\
23 Mar 22 & community & quarantine  & \subr{GenZedong} & Reddit
& 3,677 & 226,832 & 2,176  & 39,053  & -- & -- & -- & -- \\
7 Nov 17 & community & ban  & \subr{Incels} & Reddit 
& 2,862 & 273,009 & \xmark\  & \xmark\  & 2,551 & 841,379 & 2,008 & 563,354 \\
29 Jun 20 & community & ban  & \subr{The\_Donald} & Reddit
& 2,797 & 221,682 & \xmark\  & \xmark\  & 2,364 & 680,784 & 1,948 & 488,109 \\
10 Sep 18 & community & ban  & \subr{MillionDollarExtreme} & Reddit
& 2,121 & 129,276 & \xmark\  & \xmark\  & 2,049 & 536,408 & 1,762 & 401,687 \\
28 Sep 18 & community & quarantine  & \subr{TheRedPill} & Reddit
& 1,209 & 36,168 & 674 & 11,778 & --  & --  & --  & --  \\
10 Jun 15 & community & ban \& migration  & \subr{fatpeoplehate} & Reddit $\rightarrow$ Voat
& 991 & 111,751 & \xmark\  & \xmark\  & --  & -- & 991 & 39,427 \\
24 Mar 22 & community & ban  & \subr{Chodi} & Reddit
& 956 & 43,547 & \xmark\  & \xmark\  & 942 & 631,445 & 760 & 544,604 \\
21 Mar 18 & community & ban & \subr{DarkNetMarkets}  & Reddit
& 913 & 32,712 & \xmark\  & \xmark\  & 756 & 119,036 & 394 & 16,348 \\
29 Jun 20 & community & ban  & \subr{ConsumeProduct} & Reddit
& 725 & 26,786 & \xmark\  & \xmark\  & 660 & 207,087 & 501 & 138,251 \\
29 Jun 20 & community & ban  & \subr{GenderCritical} & Reddit
& 685 & 40,976 & \xmark\  & \xmark\  & 570 & 94,239 & 410 & 72,306 \\
23 Feb 23 & community & quarantine  & \subr{goblin} & Reddit
& 608 & 20,840 & 402 & 8,995 & --  & --  & --  & -- \\
14 Mar 18 & community & ban & \subr{SanctionedSuicide} & Reddit 
& 583 & 27,002 & \xmark\  & \xmark\  & 450 & 142,493 & 311 & 52,442 \\
29 Jun 20 & community & ban  & \subr{DarkHumorAndMemes} & Reddit
& 559 & 15,594 & \xmark\  & \xmark\  & 533 & 185,567 & 472 & 139,739 \\
12 Sep 18 & community & ban \& migration & \subr{greatawakening} & Reddit $\rightarrow$ Voat  & 533 & 56,780 & \xmark\  & \xmark\ & --  & --  & 533  & 16,344 \\
15 Aug 17 & community & ban  & \subr{Physical\_Removal} & Reddit
& 328 & 12,921 & \xmark\  & \xmark\  & 325 & 370,492 & 269 & 222,245 \\
29 Jun 20 & community & ban  & \subr{DebateAltRight} & Reddit
& 250 & 12,093 & \xmark\  & \xmark\  & 218 & 53,216 & 165 & 33,548 \\
21 Mar 18 & community & ban  & \subr{GunsForSale} & Reddit
& 235 & 5,837& \xmark\  & \xmark\  & 231 & 33,758 & 193 & 5,222 \\
29 Jun 20 & community & ban  & \subr{ShitNeoconsSay} & Reddit
& 160 & 4,592 & \xmark\  & \xmark\  & 142 & 39,813 & 87 & 25,351 \\
\bottomrule
\end{tabular}}
\caption{Moderation interventions included in \data. For each intervention, we report its date of application, the target and type, the platform and communities affected, as well as the number of users ({\small\faUserFriends}) and comments ({\small\faComments}) made available across the four data slices (\texttt{IN}/\texttt{OUT} and \texttt{BEFORE}/\texttt{AFTER}). The cross symbol (\xmark) indicates structurally unavailable data in the \texttt{IN-AFTER} slice. In such cases, \texttt{OUT} data is provided instead. Table rows are ordered by descending number of affected users ($\downarrow$).}
\label{tab:dataset}
\end{table*}
 
\section{Dataset Description}
Table~\ref{tab:dataset} provides a comprehensive overview of \data. The left set of columns summarizes key attributes of each moderation intervention, including its date of application, target, type, moderated space, and platform. The right set of columns details the data available for that intervention across the four data slices defined by our model.

\subsection{Available Data Slices per Intervention}
The \texttt{IN-BEFORE} slice is available for all interventions, whereas the availability of the remaining slices depends on the intervention type, as discussed in the following.

\textbf{Post removals and community quarantines.} For interventions that do not completely shut a community, we are able to provide both \texttt{IN-BEFORE} and \texttt{IN-AFTER} data. This situation represents a convenient setting for analyzing the effects of moderation within the affected space. In \data, interventions of this kind are post removals and community quarantines. Post removals correspond to the deletion of individual items of content~\cite{jhaver2024bystanders}, while community quarantines on Reddit restrict a community's visibility and accessibility without removing it, typically by adding warning interstitials and limiting discovery~\cite{chandrasekharan2022quarantined}. Because activity within the moderated space remains observable after these interventions, \data does not include \texttt{OUT} data slices for this group.

\textbf{Community bans.} After community bans, the moderated space becomes inaccessible and no post-intervention activity within the community can occur. Consequently, \texttt{IN-AFTER} data are structurally unavailable. In these cases, for each user active in the \texttt{IN-BEFORE} slice, we collect their activity across the rest of the platform during the three months preceding and following the intervention. These data form the \texttt{OUT-BEFORE} and \texttt{OUT-AFTER} slices. This design enables the study of moderation effects beyond the banned space, possibly capturing within-platform spillover dynamics~\cite{trujillo2022make}.

\textbf{Migrations following community bans.} For a subset of high-profile community bans, a considerable number of affected users migrated to the alternative Reddit-like platform Voat, where they re-created the banned communities. Because many users retained the same usernames across platforms, this setting enables direct observation of behavioral changes associated with cross-platform migration following a ban~\cite{monti2023online}. In these cases, \data includes \texttt{IN-BEFORE} data capturing activity within the later-banned Reddit community prior to the intervention, and \texttt{OUT-AFTER} data capturing activity in the newly created Voat community after the ban. For this type of intervention, Table~\ref{tab:dataset} reports the same number of users in the \texttt{IN-BEFORE} and \texttt{OUT-AFTER} slices, as only users with a matched username across platforms are included. This configuration supports analyses of moderation effects that unfold through platform migration rather than within the original moderated space and platform~\cite{horta2021platform}.

Finally, \data includes some noteworthy cases in which the same community is affected by multiple moderation interventions and therefore appears in more than one row of Table~\ref{tab:dataset}. In detail, \subr{The\_Donald} and \subr{ChapoTrapHouse} both received a quarantine followed by a definitive ban~\cite{trujillo2022make,shen2022tale}. Sequences of interventions applied to the same space enable the study of cumulative, compound, or adaptive effects of moderation over time, as well as how communities and users respond to repeated enforcement actions. In addition, for \subr{greatawakening} we provide two distinct post-ban configurations: \textit{(i)} one capturing post-intervention activity of affected users on the original platform, and \textit{(ii)} another capturing activity following migration to Voat. As a result, \subr{greatawakening}'s ban is represented twice in \data, with different \texttt{OUT-AFTER} slices. In the migration case, the corresponding \texttt{IN-BEFORE} slice is also adjusted to include only users who later migrated, conveniently enabling analyses of cross-platform behavioral change.

\begin{figure}[t]
    \centering
    \includegraphics[width=\columnwidth]{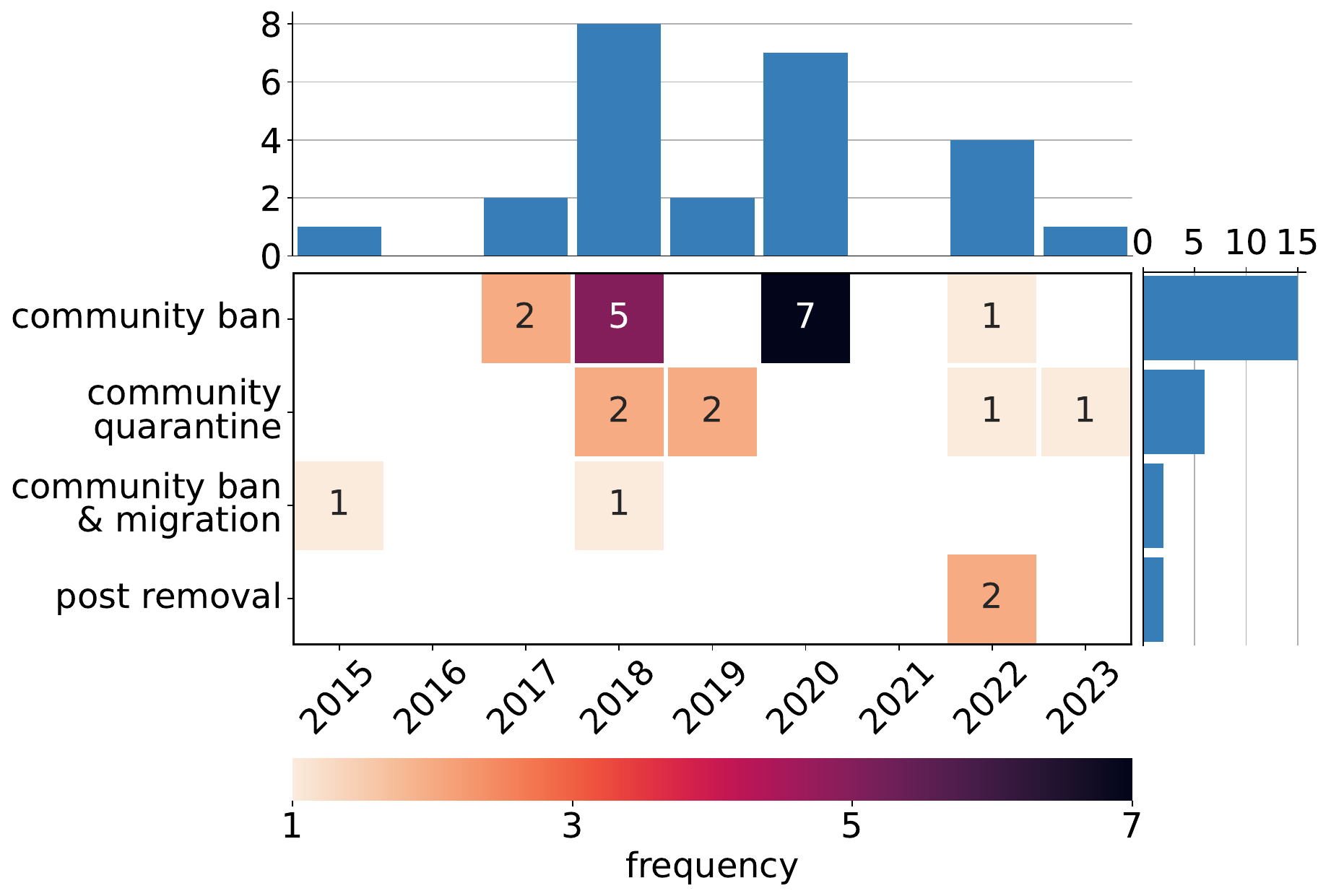}
    \caption{Frequency of interventions in the \data dataset, by type (\textit{y} axis) and year (\textit{x} axis) (central heatmap), with marginal distributions (top and right bar charts).}
    \label{fig:year-int}
\end{figure}

\subsection{Dataset Overview and Descriptive Statistics}
The data reported in Table~\ref{tab:dataset} show that \data exhibits substantial heterogeneity in the scale of the moderated communities and the corresponding affected activity. The number of affected users ranges from a few hundred to approximately 274K, while the volume of comments spans from a few thousand to over 11 million. This variation reflects the diversity of the targeted spaces, from large, mainstream subreddits such as \subr{Ask\_Reddit} and \subr{science} to smaller, fringe communities such as \subr{ShitNeoconsSay} and \subr{DebateAltRight}. These communities also span a wide range of thematic domains, ranging from general-purpose or informational communities (e.g., \subr{Ask\_Reddit}, \subr{science}), to political and ideological spaces (e.g., \subr{The\_Donald}, \subr{ChapoTrapHouse}, \subr{GenZedong}, \subr{Chodi}), to fringe or problematic communities centered on hostility, exclusion, or harassment (e.g., \subr{Incels}, \subr{TheRedPill}, \subr{fatpeoplehate}, \subr{DebateAltRight}), as well as communities associated with harmful or high-risk topics such as self-harm, extremism, illicit activity, or violence (e.g., \subr{SanctionedSuicide}, \subr{DarkNetMarkets}, \subr{GunsForSale}). Importantly, the underlying motivations for moderation also vary. Some interventions arise from violations of community-specific norms and are typically enforced by volunteer moderators of some specific subreddits, whereas others result from clear breaches of platform-wide policies and are imposed by Reddit staff. The latter frequently occur in response to illegal activity, toxic behavior, harassment, hate speech, or incitement to violence. Together, these differences highlight the heterogeneity and complexity of the content moderation landscape, a relevant portion of which is captured within \data.

\begin{figure*}[t]
    \centering
    \includegraphics[width=\textwidth]{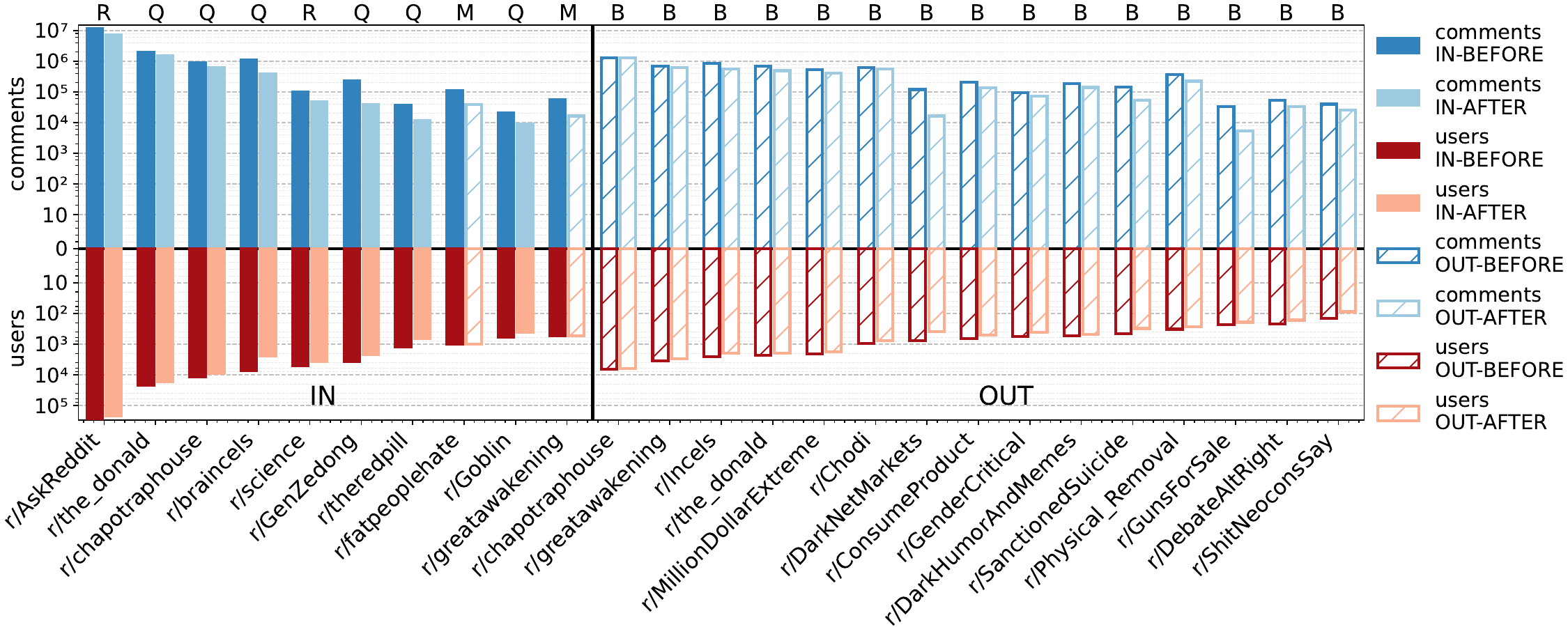}
    \caption{Number of comments (top panels) and active users (bottom panels) before and after each moderation intervention. For each intervention, bars compare pre- and post-intervention activity. The left panels show comparisons within the moderated space for interventions where \texttt{IN-AFTER} data are available. In case of bans with migrations (M), \texttt{IN-BEFORE} data is compared to \texttt{OUT-AFTER} data. The right panels show comparisons for community bans (B) using \texttt{OUT-BEFORE} and \texttt{OUT-AFTER} data. The \textit{y} axis is on a logarithmic scale to accommodate the large variation in activity volumes across communities. Subreddits on the \textit{x} axis are ordered from left to right by decreasing number of affected users.}
    \label{fig:users-comments}
\end{figure*}

Figure~\ref{fig:year-int} further illustrates heterogeneity in the temporal and typological distribution of interventions included in \data. 
Community bans and quarantines are the most frequent interventions, accounting for 15 (62.5\% of all interventions) and 6 (24\%) cases, respectively. Migrations and post removals are less represented, mirroring their lower frequency both in practice and in the existing literature~\cite{horta2021platform,jhaver2024bystanders}. The year with the highest concentration of interventions in \data is 2018, which includes five bans, two quarantines, and one ban followed by migration, for a total of eight interventions (32\%). Overall, the dataset combines well-studied, high-profile interventions with more recent cases that have received little or no prior attention, such as the quarantines of \subr{goblin} (2023) and \subr{GenZedong} (2022), and the ban of \subr{Chodi} (2022). Notably, intervention frequency does not directly correspond to impact. For instance, the two post removals in \data affect some of the largest user groups, with approximately 274K and 5K users, respectively.

Figure~\ref{fig:users-comments} summarizes changes in activity associated with each intervention by comparing pre- and post-intervention volumes of comments (top) and active users (bottom). For interventions with post-intervention activity---community quarantines (Q), bans with migrations (M), and post removals (R)---the figure compares \texttt{IN-BEFORE} and \texttt{IN-AFTER}, except for the two migration bans, where \texttt{IN-BEFORE} is compared to \texttt{OUT-AFTER}.
For community bans (B), the right panel reports comparisons between \texttt{OUT-BEFORE} and \texttt{OUT-AFTER} activity. Across nearly all interventions, both the number of comments and the number of active users decrease following moderation, regardless of intervention type or community size, though the magnitude of the change varies substantially across cases. These descriptive patterns are consistent with prior findings reported for specific interventions~\cite{trujillo2022make} and suggest that moderation may be generally associated with reduced user activity and participation across a broad range of contexts.
 \section{Use Cases}
We envision \data to support a wide range of research tasks in the study of online content moderation, as outlined below. Across all use cases and tasks, the dataset enables researchers to develop and validate methods consistently across multiple interventions, supporting large-scale and comparative analyses that are difficult to conduct with the existing fragmented resources. The listed use cases are not exhaustive, and we expect the dataset to enable additional lines of inquiry beyond those discussed here.

\textbf{Assessing effects and effectiveness.} \data enables systematic analyses of how moderation interventions affect user~\cite{horta2025deplatforming} and community behavior~\cite{chandrasekharan2022quarantined}, supporting both descriptive and quasi-experimental evaluations of intervention outcomes. By providing aligned pre- and post-intervention data, scholars can measure short- and medium-term changes in activity, engagement, participation patterns~\cite{trujillo2022make}, and other behavioral indicators~\cite{tessa2025quantifying}, and compare the effectiveness of different intervention types. The dataset further supports analyses of heterogeneity, allowing effects to be examined across users, communities, platforms, and intervention characteristics. Furthermore, because the dataset spans interventions from 2015 to 2023 and includes multiple instances of each intervention type, it enables longitudinal analyses of moderation practices, making it possible to assess whether the effects of similar interventions remain stable over time or evolve alongside platforms, policies, and user behavior.

\textbf{Assessing fairness, consistency, and equity.} \data supports research on the fairness and consistency of moderation practices by enabling comparisons across users, communities, platforms, and intervention instances. For example, by analyzing \texttt{IN-BEFORE} activity, researchers can examine the behavioral and content-level signals that precede moderation~\cite{habib2022proactive}, shedding light on potential triggers and decision patterns underlying enforcement actions. Extending this analysis across multiple instances of the same intervention type on the same platform makes it possible to assess whether similar behaviors are treated consistently over time or across contexts, and to identify systematic differences in how moderation is applied. In turn, such analyses can reveal disparities in moderation enforcement or its outcomes, and inform studies of bias, equity, and accountability~\cite{lee2019procedural}.

\textbf{Assessing the role of context.} By spanning multiple intervention types, platforms, and communities, \data enables comparative analyses that examine how contextual factors shape both moderation decisions and their outcomes~\cite{shen2022tale,chandrasekharan2022quarantined,cresci2022personalized}. Researchers can study how characteristics such as community size and norms, topical focus, and participation habits mediate the application and effects of moderation interventions, and whether similar interventions yield different outcomes across contexts. This contextual perspective helps identify regularities and divergences in moderation practices, disentangling effects driven by the intervention from those shaped by social and organizational context.

\textbf{Predictive modeling and early-warning systems.} Beyond retrospective analyses, \data supports predictive approaches that aim to anticipate the outcomes of moderation interventions before they occur. By using pre-intervention signals as inputs and post-intervention outcomes as ground-truth, scholars can train models to forecast user-level responses such as disengagement, migration, or behavioral change~\cite{tessa2025beyond}, as well as community-level effects including shifts in activity, toxicity, or participation diversity~\cite{tessa2025quantifying}. Prior work in this area has explored such tasks in a limited number of settings, typically focusing on a single intervention type or platform~\cite{niverthi2022characterizing,habib2022proactive}. However, the breadth and longitudinal coverage of \data enable large-scale evaluations of predictive models, including assessments of their robustness and generalization across time, intervention types, and platforms. This, in turn, opens the door to the development and validation of early-warning systems that can inform more proactive and context-aware moderation strategies.

\textbf{Empowering transparency, accountability, and governance.} By linking moderation interventions to rich behavioral context before and after enforcement, \data also supports studies on transparency, accountability, and regulatory compliance by complementing existing transparency initiatives that primarily provide metadata about moderation decisions but lack information about user activity and responses~\cite{trujillo2025dsa,kaushal2024automated}. For example, researchers can integrate this dataset with external data sources such as transparency disclosures and platform policies to assess the consistency between actual moderation actions and the declared enforcement practices or platform rules in place at the time of the intervention~\cite{tessa2025improving}. To this regard, \data can inform both scholarly and policy-oriented investigations into how moderation interventions shape online ecosystems and how governance mechanisms operate at scale.

\textbf{Methodological development.} While the preceding use cases focus on substantively oriented questions about moderation outcomes, \data also supports methodological research on how such effects are measured. The intervention-centered structure and standardized pre- and post-intervention windows enable systematic comparisons of alternative analytical approaches, including simple before–after designs~\cite{cima2025investigating}, interrupted time series analyses~\cite{habib2022exploring,broniatowski2025explaining}, and more complex counterfactual modeling strategies~\cite{slaughter2025community,trujillo2022make}. Researchers can use the dataset to study the sensitivity of results to modeling choices such as window length, temporal aggregation, outcome definition, or the inclusion of external activity as a comparison signal.By offering a shared empirical basis with multiple interventions across time and contexts, the dataset supports benchmarking, robustness checks, and methodological validation, fostering more reliable and transparent approaches to measuring content moderation effects.

\textbf{Extensions for causal analyses.} Finally, \data can serve as a foundation for more robust causal analyses by supporting the construction of external control groups, as done in some recent works~\cite{seckin2025labeled}. For each moderated intervention and space, interested scholars could identify independent but comparable and unmoderated spaces---e.g., matched on characteristics such as topic, size, or user composition---and use them as controls within quasi-experimental designs~\cite{cima2025investigating}.
While not included in the current TBBT release, the dataset’s intervention-centered structure supports these extensions and can strengthen causal claims, enabling more rigorous assessment of whether observed effects are attributable to moderation interventions.
 \section{Limitations and Future Work}
\data is designed to support a broad range of analyses, but faces some limitations. These also point to valuable directions for future work. First, the dataset exhibits a platform and intervention skew, with a predominant focus on Reddit and on community bans and quarantines. This reflects both constraints on data access and the emphasis of existing research. As such, future extensions of the dataset could incorporate additional platforms and a wider spectrum of interventions, particularly softer or hybrid moderation mechanisms. Second, the present version of \data is observational in nature, which limits the extent to which causal claims can be drawn directly from this data. While the intervention-centered design supports quasi-experimental analyses, robust causal estimates require the construction of appropriate control groups or counterfactuals.
As discussed in the use cases, the dataset supports such extensions, and we encourage researchers to augment it with matched control spaces or external comparison groups.
Then, for certain interventions (e.g., community bans), post-intervention activity within the moderated space (\texttt{IN-AFTER}) is structurally unavailable, as the space itself no longer exists after the moderation. This is an inherent property of deplatforming interventions rather than a limitation of our data collection process. Nonetheless, to mitigate this limitation, \data explicitly includes activity by affected users outside the moderated space, enabling analyses of post-intervention behavior, spillover effects, and migrations. Furthermore, the dataset introduces a degree of user selection bias by focusing on users with a minimum level of pre-intervention activity. This choice reduces noise by focusing on those users meaningfully exposed to the moderation, but it may underrepresent low-activity users.
Additionally, the bot filtering step of our data preparation pipeline may not be exhaustive, meaning that some bots might be undetected. Finally, \data spans interventions from 2015 to 2023---a period during which platforms, policies, and user norms have evolved substantially. This temporal drift complicates direct comparisons across interventions conducted at different times. However, it also offers opportunities to study how moderation practices and their effects changed over time, providing a longitudinal perspective on the evolution of content moderation.
 \section{Conclusions}
Research on online content moderation is constrained by limited data access and fragmented resources. To mitigate these issues, we introduced \textit{The Big Ban Theory} (\data)---an intervention-centered dataset that aggregates pre- and post-moderation activity for a large set of moderation interventions across multiple years, platforms, and intervention types. By structuring data along orthogonal space and time dimensions, and explicitly accounting for cases where post-intervention data within the moderated space is unavailable, \data supports a broad range of descriptive, methodological, and predictive analyses. While the dataset has certain clear limitations, it provides a structured and extensible foundation for studying moderation interventions at scale. This resource will facilitate cumulative research on content moderation and support ongoing efforts to understand, evaluate, and improve moderation practices in online platforms. 
\section{Ethical Statement}
Data included in \data originate from publicly available content on the source platforms. To mitigate privacy risks, we pseudonymized all user, content, and community identifiers, reducing risks of re-identification while preserving internal consistency for downstream analysis. Nevertheless, as with any dataset derived from online activity, residual re-identification risks remain, and the data should be handled accordingly. Moderation interventions can be socially and personally sensitive, and \data reflects interactions within communities and by users that were subject to moderation. As a consequence, the data may contain offensive, harmful, or otherwise triggering content. The dataset does not encode normative judgments about the appropriateness of moderation decisions, and analyses should be interpreted with care to avoid stigmatizing individuals or communities. Moreover, findings derived from this dataset should not be used to automate or justify punitive decisions without appropriate human oversight, as misuse could lead to unintended or harmful societal impacts. Finally, \data is released in accordance with FAIR principles and is intended exclusively for scientific research. Access and use are governed by terms of use that prohibit attempts at re-identification, or uses that could harm individuals or communities represented in the data. We expect researchers to adhere to these conditions and to established community norms for ethical research on online platforms.
 \section{Acknowledgments}
This work is partly supported by the ERC project DEDUCE under grant \#101113826, and by the European Union -- Next Generation EU, Mission 4 Component 1, for project PIANO (CUP B53D23013290006).

\bibliography{references}

\newpage
\begin{table*}[t]
\centering
\adjustbox{max width=\textwidth}{
    \begin{tabular}{llc}
    \toprule
    \textbf{data field} & \textbf{description} & \textbf{pseudonymized} \\
    \midrule
    \texttt{comment\_id}                     & Comment ID & \cmark \\
    \texttt{created\_utc}           & Date and time (UNIX timestamp) on which the comment was created &  \\
    \texttt{edited}                 & Whether (and when) a comment has been edited after being created &  \\
    \texttt{stickied}               & Boolean indicating whether the comment has been stickied by the moderators &  \\
    \texttt{selftext}               & Full text content of the submission/comment & \cmark \\
    \texttt{subreddit}              & The name of the subreddit (or subverse) the comment was made in &  \\
    \texttt{link\_id}             & ID of the submission the comment belongs to & \cmark \\
    \texttt{author}                 & Username of comment author & \cmark \\
    \texttt{author\_created\_utc}   & Creation time of author account (UNIX timestamp) &  \\
    \texttt{author\_flair\_richtext}& Content of the author’s flair &  \\
    \texttt{score}                  & Overall score (upvotes minus downvotes) &  \\
    \texttt{all\_awardings}         & List of awardings added to the comment &  \\
    \texttt{gilded}                 & The number of gild awards a comment has received &  \\
    \texttt{gildings}               & Details of gild awards a comment has received &  \\
    \bottomrule
    \end{tabular}}
    \caption{Standardized data fields included in \data, their brief description, and a flag indicating whether each field has been pseudonymized. In the \texttt{self\_text} field, only user mentions have been pseudonymized.}
\label{tab:attributes-description}
\end{table*}
 
\section{Appendices}
\subsection{Available Fields and Pseudonymization}
Table~\ref{tab:attributes-description} lists all data fields included in \data, together with a brief description and an indication of whether each field has been pseudonymized. From the full set of available platform-specific fields, we selected a subset that balances informational richness with consistency across interventions, platforms, and nearly a decade of data collection. This process involved harmonizing field names and semantics to align data originating from different platforms and evolving data formats. All fields that may possibly contain identifiable information---such as user identifiers, content identifiers, and usernames---were processed using a deterministic cryptographic hash function, ensuring consistent pseudonyms across the dataset. This pseudonymization was applied to both metadata fields (e.g., \texttt{id}, \texttt{created\_at}) and selected textual fields (e.g., \texttt{author}). 
Notably, in the \texttt{self\_text} field, we pseudonymized only user mentions (e.g., \texttt{u/username}), ensuring consistency with the \texttt{author} field.

\subsection{Post Removal Data Considerations}  \label{sec:postremoval}
Removed posts are not included in any data slices for the post removal intervention, as they are not accessible. Instead, we rely on the timing of removal activity affecting specific subreddits, as reported by \citet{jhaver2024bystanders}. In particular, they report a peak in post removals on Reddit in June 2022, referred to as the treatment period. Therefore, we collected comments from the affected subreddits three months before and after the reference date \textit{$t_0$}, corresponding to the one-month treatment period, in order to capture users' reactions and the discussions surrounding the intervention. 
\end{document}